\documentclass[12pt]{article}
\usepackage{epsfig,graphicx,amssymb,amsfonts}

\def \be {\begin{equation}}
\def \ee {\end{equation}}
\def \bea {\begin{eqnarray}}
\def \eea {\end{eqnarray}}
\def \nn {\nonumber}

\def \R {\mathbb{R}}

\def \H {\mathbb{H}}

\def \M {\mathbb{M}}

\def \dels {\partial\kern-.5em / \kern.5em}
\def \As {{A\kern-.5em / \kern.5em}}
\def \Ds {D\kern-.7em / \kern.5em}

\def \a {\alpha}
\def \b {\beta}

\def \d {\delta}
\def \eps {\epsilon}

\def \s {\epsilon}

0mm \makeatletter \@addtoreset{equation}{section}

\makeatother

\newcommand{\hs}[1]{\hspace{#1 mm}}
\renewcommand{\a}{\alpha}
\renewcommand{\b}{\beta}
\renewcommand{\c}{\gamma}
\renewcommand{\d}{\delta}
\newcommand{\e}{\epsilon}

\newcommand{\vp}{\varphi}
\def\bbox{{\,\lower0.9pt\vbox{\hrule \hbox{\vrule height 0.2 cm
\hskip 0.2 cm \vrule height 0.2 cm}\hrule}\,}}
\newcommand{\dsl}{\pa \kern-0.5em /}

\newcommand{\pa}{\partial}
\newcommand{\p}[1]{(\ref{#1})}

\def \K {{\tt I\kern-.25em K}}

\begin{document}


\begin{titlepage}
\begin{center}

\hfill\parbox{4cm}{
{\normalsize\tt hep-th/0306291}\\
{\normalsize OU-HET 449} }

\vskip .5in

{\LARGE \bf Hyperbolic Space Cosmologies}

\vskip 0.5in

{\large %
Chiang-Mei Chen$^a$\footnote{{\tt cmchen@phys.ntu.edu.tw}}, %
Pei-Ming Ho$^a$\footnote{{\tt pmho@phys.ntu.edu.tw}}, %
Ishwaree P. Neupane$^a$\footnote{{\tt ishwaree@phys.ntu.edu.tw}},
\\
Nobuyoshi Ohta$^b$\footnote{{\tt ohta@phys.sci.osaka-u.ac.jp}}
and John E.\ Wang$^{a}$\footnote{{\tt hllywd2@phys.ntu.edu.tw}} }

\vskip 0.5in

${}^a$ {\it Department of Physics, National Taiwan University,
Taipei 106, Taiwan}\\[3pt]
${}^b$ {\it Department of Physics, Osaka University,
 Toyonaka, Osaka 560-0043, Japan}\\
[0.3in]

{\normalsize June 2003}

\end{center}

\vskip .3in

\begin{abstract}
\normalsize\noindent We present a systematic study of accelerating
cosmologies obtained from M/string theory compactifications of
hyperbolic spaces with time-varying volume. A set of vacuum
solutions where the internal space is a product of hyperbolic
manifolds is found to give qualitatively the same accelerating
four-dimensional FLRW universe behavior as a single hyperbolic
space. We also examine the possibility that our universe is a
hyperbolic space and provide exact Milne type solutions, as well
as intersecting S-brane solutions. When both the usual $4D$
spacetime and the $m$-dimensional internal space are hyperbolic,
we find eternally accelerating cosmologies for $m\geq 7$, with and
without form field backgrounds. In particular, the effective
potential for a magnetic field background in the large 3
dimensions is positive definite with a local minimum and thus
enhances the eternally accelerating expansion.

\end{abstract}

\vfill

\end{titlepage}
\setcounter{footnote}{0}

\pagebreak
\renewcommand{\thepage}{\arabic{page}}
{\baselineskip=5mm\tableofcontents}


\setcounter{footnote}{0} \setcounter{page}{2}

\section{Introduction}

The past few years have produced a surge in the study of
time-dependent solutions due to experimental data which point to
cosmic acceleration in the present universe.  While many effective
inflationary scenarios may be devised to explain this phenomenon,
one would hope that a natural candidate would emerge by studying a
more fundamental theory such as superstring or M-theory.  Until
recently, however, it was believed that the low energy
supergravity limit of string theory could not generate four
dimensional acceleration from compactification.  This
argument~\cite{NM}, often regarded as a no-go theorem, is valid
when the compact internal space is a time independent,
non-singular compact manifold without boundary.

By explicit construction, Townsend and Wolhfarth showed in
Ref.~\cite{TW} that vacuum solutions can produce accelerating
cosmologies if one uses compact time dependent hyperbolic internal
spaces. Their model was recognized~\cite{NO2} as the zero flux
limit of the S2-brane solutions~\cite{CM,NO1}, and the cosmology
of these generalized S-brane solutions has also been studied.
Further generalizations of these models were discussed in
Refs.~\cite{Roy}-\cite{MI}.
Related cosmologies have been discussed in
Refs.~\cite{LMP}-\cite{GPCZ}. The use of hyperbolic space for
internal space was proposed in Refs.~\cite{kaloper,silva}.

This class of S-brane solutions provides a way to obtain cosmic
acceleration from time dependent compactification where the
internal dimensions are hyperbolic and the usual $(3+1)$
dimensions are flat. In this and related scenarios, the typical
amount of inflation is on the order of one, which is too small to
solve the horizon and flatness problems in the early
universe~\cite{Wolf,EG,NO3,CHNW}. Consequently it is of utmost
importance to try to find a suitable mechanism to generate a
larger inflationary period.

One might suspect that since the effective potentials due to
compactifications are always exponentials with coefficients of
order one, the e-folding number within the acceleration phase
should also be of order one. But it was pointed out in
Ref.~\cite{CHNW} that an exponential potential with suitable
coefficients of scalar fields can in principle produce eternal
inflation. In this paper we make a further step by presenting
solutions of eternally accelerating expansion for pure gravity
compactified on hyperbolic spaces.

In our search for more general models with sufficient inflation,
we have tried to make a systematic study of the models
compactified on a product of hyperbolic and flat spaces.  For
general product space compactifications, solutions are generally
difficult to obtain.  Using a specific ansatz we find a new class
of vacuum spacetimes which are a product of flat and hyperbolic
spaces.  We examine the possibility of using these spaces to
obtain inflation but find that they do not lead to accelerating
cosmologies when the external dimensions are flat.

We next turn to the prospect of having hyperbolic external
dimensions.  We first study the case where the external space is
hyperbolic and the internal space is flat.  As a further step we
study the case where both the external and internal dimensions are
hyperbolic since these spaces seem to be among the most promising
for obtaining large inflation. In this case we find solutions
whose late time behavior is approximately characterized as a Milne
spacetime with nearly constant expansion. By studying perturbative
expansions about such Milne solutions, we find that it is possible
to obtain eternally accelerating expansion when the dimension of
the internal space is greater than or equal to seven, which is
interesting in view of the possibility of embedding these
solutions in M-theory.  While it is not clear if this model is
phenomenologically viable, at least we find that it is possible to
improve the amount of inflation for M-theory compactifications.

This paper is organized as follows: In section~2, we discuss
vacuum solutions to Einstein equations which are products of flat
and hyperbolic spaces and show that they lead to a phase of
accelerated expansion. Here we also review and further discuss why
hyperbolic internal spaces help inflation in general. In
section~3, we discuss the differences and advantages in treating
hyperbolic spaces instead of flat spaces as the external space. In
addition we find Milne type solutions as exactly solvable
examples.  In section~4, we discuss the product spaces by using
the effective $4D$ action with scalar fields which parametrize the
radii of the internal spaces. This viewpoint is useful because we
can use our knowledge of scalar field inflation to increase our
physical intuition of our results. The effective action approach
is used in section~5 to analyze Milne type exact solutions for the
case in which both the internal and external spaces are
hyperbolic. Perturbations around Milne spacetimes are shown to
lead to eternally accelerating expansion when the dimension of the
internal space is greater than or equal to seven. Our result
suggests that there is the possibility of obtaining big expansion
from M-theory compactifications.

\section{Product spaces: gravity viewpoint}\label{Sols}

\subsection{Vacuum solutions for higher-dimensional gravity}
\label{higherdimgrav}
In this section, we discuss and construct vacuum solutions of the
Einstein equations.  We focus primarily on solutions which can be
written as a product of hyperbolic and flat spaces.  In the case
of these vacuum solutions, accelerated expansion is possible only
if the internal dimensions include hyperbolic spaces, i.e. have
negative curvature.

Let us consider a general spacetime with the following product
space metric ansatz
\begin{equation}
\label{metric} ds_D^2 = - {\rm e}^{2A(t)} dt^2 + \sum_{i=0}^n
{\rm e}^{2B_i(t)} d\Sigma^2_{m_i,\epsilon_i},
\end{equation}
where the $\Sigma_{m_i,\epsilon_i}$ are $m_i$ dimensional spaces
with curvature specified by $\epsilon_i$; the values of
$\epsilon_i=0,+1,-1$ correspond to the flat, spherical or
hyperbolic spaces, respectively.  The metric for each $\Sigma$ is
\begin{equation}
d\Sigma_{m_i,\epsilon_i}^2 = \bar g^{(i)}_{ab} dz^a dz^b = \left\{
\begin{array}{ll}
 d \psi^2 + \sinh^2\psi \, d\Omega_{m_i-1}^2, \qquad & \epsilon_i=-1,\\
 d \psi^2 + \psi^2 \, d\Omega_{m_i-1}^2, \qquad & \epsilon_i=0,\\
 d \psi^2 + \sin^2\psi \, d\Omega_{m_i-1}^2, \qquad & \epsilon_i=+1,
\end{array} \right.
\label{gmetric}
\end{equation}
which have curvature
\begin{equation}
\bar R^{(i)}_{ab} = \epsilon_i (m_i - 1) \bar g^{(i)}_{ab}.
\end{equation}
We will also denote the metrics in eq.~(\ref{gmetric}) for
$m$-dimensional flat, spherical and hyperbolic spaces by
$ds^2_{Rm}$, $ds^2_{Sm}$ and $ds^2_{Hm}$. The Ricci tensor for the
metric (\ref{metric}) is relatively simple
\begin{eqnarray}
R_{tt} &=& - \sum_{i=0}^n m_i (\ddot B_i + \dot B_i^2 - \dot A
\dot B_i),
\\
R^{(i)}_{ab} &=& \left\{ {\rm e}^{2B_i-2A} \left[ \ddot B_i + \dot
B_i \left( - \dot A + \sum_{j=0}^n m_j \dot B_j \right) \right] +
\epsilon_i (m_i - 1) \right\} \, \bar g^{(i)}_{ab}.
\end{eqnarray}

To simplify the equations we use the gauge condition
\begin{equation}
\label{gauge} - A + \sum_{j=0}^n m_j B_j = 0\,.
\end{equation}
The vacuum Einstein equations then become
\begin{eqnarray}
- \, \sum_{i=0}^n m_i \ddot B_i + \sum_{i=0}^n m_i (m_i-1) \dot
B_i^2 + \sum_{j\neq i}^n m_i m_j \dot B_i \dot B_j &=& 0 \,,
\label{EqCon}
\\
\ddot B_i + \eps_i (m_i - 1) {\rm e}^{2(m_i-1)B_i} \, {\rm
e}^{2\sum_{j\neq i}m_j B_j} &=& 0 \, \label{EqBi}.
\end{eqnarray}
For spacetimes given in eq.~\p{metric}, the $D$-dimensional
Einstein frame metric is defined as
\begin{eqnarray}
ds_D^2 = \sum_{i=1}^n {\rm e}^{2B_i(t)} d\Sigma_{m_i,\e_i}^2 +
{\rm e}^{-\frac{2}{d-1}\sum_{i=1}^n m_i B_i(t)} ds_{E,d+1}^2,
\label{einframe}
\end{eqnarray}
where $d\equiv m_0$, and the ($d+1$)-dimensional part of the
metric is given by
\begin{equation} \label{EinM}
ds_{E,d+1}^2 = {\rm e}^{\frac2{d-1}\sum_{i=1}^n m_i B_i} \left( -
{\rm e}^{2\sum_{j=0}^n m_j B_j} dt^2 + {\rm e}^{2B_0} ds_{d}^2
\right).
\end{equation}

\subsection{Product space of same type subspaces}

Due to the coupling inherent in the equations of motion for spaces
with non-zero curvature, it is not known how to generally solve
the coupled differential equations in eqs.~\p{EqCon} and \p{EqBi}.
We can find exact solutions only in some relatively simple cases
if we take a particular ansatz for the solution. As an interesting
case that can be solved exactly, we consider the spacetime with a
$d$-dimensional flat subspace, so $\e_0=0$.  For simplicity we
assume that the rest of space is a product of spaces with the same
curvature $\epsilon_1=\cdots=\epsilon_n=\epsilon$. These spaces
are of the form $\R^{1+d} \times \M_{m_1,\epsilon} \times \cdots
\times \M_{m_n,\epsilon}$.

In this case, the equation for $B_0$ is easily solved
\begin{equation} \label{B0}
B_0 = \lambda_0 t + \lambda_1,
\end{equation}
with two integration constants $\lambda_0$ and $\lambda_1$. The
constant $\lambda_1$ may be eliminated by a shift of the time in
$\R^{1,d}$, and in the following discussion we take $\lambda_1=0$.

For the functions $B_i$ we make the ansatz
\begin{equation}
B_i = - \frac{d\lambda_0}{m-1} t + \frac{\beta_i}{m-1} -
\frac1{m-1} f(t), \label{vs1}
\end{equation}
where we have defined $m=\sum_{i=1}^n m_i$ and
\begin{equation} \label{ft}
 f(t) = \left\{ \begin{array}{ll}
 \ln\left( \sinh[\lambda(t-t_1)] \right), \qquad & \epsilon=-1,
\\
 \lambda (t-t_1), & \epsilon=0,
\\
\ln\left( \cosh[\lambda(t-t_1)] \right), & \epsilon=+1,
\end{array} \right.
\end{equation}
and $\beta_i$ and $\lambda$ are undetermined constants. The
variable $A$ in this case is then determined by the gauge
condition~\p{gauge}, namely,
\begin{equation} \label{ft1}
A = - \frac{d\lambda_0}{m-1} t + \frac{\beta}{m-1} -
\frac{m}{m-1}f(t),
\end{equation}
with $\beta=\sum_{i=1}^n m_i \beta_i$.  For different values of
$i$, the equations in eq.~(\ref{EqBi}) determine the $\beta$
parameters
\begin{eqnarray}
\beta_i &=& \frac12 \ln \left[ \frac{\lambda^2}{(m-1)(m_i-1)}
\prod_{j=1}^n \left( \frac{m_i-1}{m_j-1} \right)^{m_j} \right],\nonumber \\
\beta &=& \frac{m}2 \ln \left[ \frac{\lambda^2}{m-1}
\prod_{i=1}^n (m_i-1)^{-m_i/m} \right],
\end{eqnarray}
while $\lambda$ is determined by eq.~(\ref{EqCon}) to be
\begin{equation} \label{lambda}
\lambda = \sqrt{\frac{d(m+d-1)}{m}} \,\lambda_0.
\end{equation}

These solutions are slight generalizations of those discussed in
Ref.~\cite{CHNW} where product spaces of identical subspaces were
considered with $m_i=m_1$ for all $i$. In this more generic case,
the ($d+1$)-dimensional metric in the Einstein frame takes the
form
\begin{equation}
\label{EinM1} ds_{d+1}^2 = - S^{2d}(t) dt^2 + S^2(t) ds_{Rd}^2,
\end{equation}
with the scale factor
\begin{equation}
S(t) = \exp\left[ \frac{-(m+d-1) \lambda_0 t + \beta - m
f(t)}{(d-1)(m-1)} \right].
\end{equation}
In terms of the proper time defined by
\begin{eqnarray}
d\tau = S^d(t) dt, \label{proper}
\end{eqnarray}
the metric~\p{EinM1} takes the standard FLRW form.

The conditions for expansion and accelerated expansion are,
respectively,
\begin{eqnarray}
\frac{dS}{d\tau} > 0, \qquad \frac{d^2 S}{d\tau^2} > 0 \,.
\label{conde}
\end{eqnarray}
For the above solution, we find that the accelerated expansion is
possible only for hyperbolic internal spaces and the
conditions~\p{conde} are
\begin{eqnarray}
n(t) \equiv -\sqrt{md} \coth\lambda(t-t_1) - \sqrt{m+d-1} &>&
0,\nonumber
\\
\frac{(m-1)d}{\sinh^2 \lambda(t-t_1)} - n^2(t) &>& 0.
\end{eqnarray}
These conditions, which depend only on $d$ and the total dimension
of the internal space, are basically the same conditions that one
obtains from a single internal space, and the expansion factor
(the ratio of the scale factors at the starting and ending times
of the accelerated expansion) is of order
one~\cite{TW,Wolf,EG,NO3,CHNW}.

Intuitively, the reason why we can get accelerated expansion only
for hyperbolic space is that the hyperbolic internal spaces act as
positive potentials in the dimensionally reduced effective
$(d+1)$-dimensional theory. We will further discuss this effective
potential viewpoint later.

\subsection{The $\M_0\times\M_1 \times\M_2$ spaces}

Our product space compactifications to an Einstein manifold
unfortunately do not give sufficient inflation as their behavior
is similar to the original model. In the following discussions we
will explore further possible inflationary mechanisms.

Consider a product of three spaces. Then the three coupled
differential equations following from~(\ref{EqBi}) must satisfy
\begin{equation}\label{coupled1}
B_0 = - \, \frac{(m_2-1)B_2+m_1 B_1}{m_0} + \frac{1}{2m_0} \,
\ln\left(\frac{-\,\ddot{B_2}}{\epsilon_2(m_2-1)}\right) \, ,
\end{equation}
\begin{equation}\label{coupled2}
\epsilon_2 (m_2-1) \, \ddot{B_1} \, {\rm e}^{2B_1} =
{\epsilon_1(m_1-1)} \, \ddot{B_2} \, {\rm e}^{2B_2} \,,
\end{equation}
\begin{eqnarray} \label{coupled3}
\frac{\partial^2}{\partial t^2}\left(\ln
\left(-\epsilon_2\ddot{B_2}\right) \right) &=& 2 m_1 \ddot{B_1} +
2 (m_2-1) \ddot{B_2} - 2 \epsilon_0 m_0 (m_0-1)\, {\rm
e}^{\frac{(m_0-1)}{m_0} \ln\left(
\frac{-\,\ddot{B_2}}{\epsilon_2(m_2-1)} \right)}
\nn \\
&{}& \qquad\qquad \times {\rm
e}^{\frac{2}{m_0}\left(m_0+m_2-1)B_2+m_1 B_1\right)}\,.
\end{eqnarray} %
It is difficult to find a solution when $\epsilon_0\neq 0$ and all
the internal spaces are non-flat. However, in some cases such as
$\R^1\times \H_3\times \H_m$, we can obtain exact solutions with
$H\propto 1/t$. This will be discussed in subsections~\ref{RHH2}
and \ref{eternal}. Here we consider the solvable case with
$\epsilon_0= 0$ and take $m_0=d=3$, but we will put no restriction
on the number of internal dimensions $m_1, m_2$.

\subsection{$\R^{3+1}\times \R^{m_1}\times \H_{m_2}$} \label{RRH}

One of the simplest examples we can solve is the case where the
internal space is the product of flat and hyperbolic spaces. We
define
\begin{equation}
B_0(t) = \lambda_0 t \,, \quad %
B_1(t) = a(t) - \frac{3\lambda_0 t}{m_1+m_2-1} \,, \quad %
B_2(t) = b(t) - \frac{3\lambda_0t}{m_1+m_2-1}\,,
\end{equation}
so that the vacuum Einstein equations (\ref{coupled1},
\ref{coupled2}, \ref{EqCon}) further simplify to
\begin{eqnarray}
&&\ddot{b} = (m_2-1) \, {\rm e}^{2m_1 a(t)+2(m_2-1)b(t)} \,,
\qquad \ddot{a} = 0 \,,\nonumber\\
&& m_1(m_1-1) {\dot a}^2 + m_2(m_2-1) {\dot b}^2 - m_2 {\ddot b} +
2 m_1 m_2 {\dot a}{\dot b} = \frac{3(m_1+m_2+2)}{(m_1+m_2-1)}
\lambda_0^2,
\end{eqnarray}
while eq.~(\ref{coupled3}) gives only a consistency condition. The
set of the above equations has the solution
\begin{equation} \label{solutionRH}
a(t)= \alpha_0 t \,, \quad %
b(t)= - \frac{m_1}{m_2-1} \, \alpha_0 t + \frac{1}{m_2-1}
\ln\left( \frac{\beta}{\sinh((m_2-1)\beta\,t)} \right) \,,
\end{equation}
with
\begin{equation} \label{betavalue}
\beta^2 = \frac{m_1(m_1+m_2-1)\alpha_0^2}{m_2(m_2-1)^2} +
\frac{3(m_1+m_2+2)}{m_2(m_2-1)(m_1+m_2-1)} \, \lambda_0^2 \,.
\end{equation}

For the case of our interest, $m_1=1, m_2=6$, the metric in
Einstein frame takes the form
\begin{equation}
ds^2 = {\rm e}^{\frac72 \lambda_0 t - a(t) - 6 b(t)} ds_{E4}^2 +
{\rm e}^{-\lambda_0t+2a(t)} \, dr^2 + {\rm e}^{-\lambda_0t+2b(t)}
ds_{H6}^2\,,
\end{equation}
where $ds_E^2$ is given by \p{EinM1} with $d=3$ and the scale
factor is
\begin{equation}
S(t) = {\rm e}^{-\frac{3\lambda_0 t}4 + \frac{a(t)}2 + 3 b(t)} \,.
\end{equation}
If we define the four-dimensional proper time $\tau$ via
\begin{equation} \label{tauandt}
d \tau = \pm \, S^3(t) dt \,,
\end{equation}
then the $4D$ spacetime is expanding if $dS/d\tau>0$, namely, if
$n(t)<0$ or $n(t)>0$, for the plus or minus sign
in~(\ref{tauandt}), where
\begin{equation}
n(t) = \frac{\alpha_0}{10} + 3\beta\,\coth(5\beta\, t) +
\frac{3\lambda_0}{4}\,.
\end{equation}
Let us write
$$
\alpha_0 = c \, \lambda_0 \,, \qquad \mbox{so}~~ \quad \beta =
\frac{|\lambda_0|}{10} \sqrt{4c^2+15}\,,
$$
where $c$ is some number. Then the condition for acceleration
$d^2S/d\tau^2>0$ is
\begin{equation}
\frac{15}{2}\frac{\beta^2}{\sinh^2(5\beta\,t)} >
\left(\frac{(2c+15)\lambda_0}{20}+3\beta
\coth\left(5\beta\,t\right)\right)^2\,.
\end{equation}
This condition is satisfied for both the positive and negative
time interval, by suitably choosing $c$. An interesting case is
$c=1/2$, so $\beta=4|\alpha_0|/5$. In this case, the time-varying
volume factor $e^{2B_1(t)}$ of the $\R^1$ becomes unity. The
condition for acceleration (as well as the condition for
expansion) is satisfied in the interval $t_1>t>t_2$ (or
$t_1<t<t_2$ depending on the choice $\lambda_0<0$ or
$\lambda_0>0$), where
\begin{equation}%
t_1 = \frac{1}{8\alpha_0} \ln\left(\frac{2-\sqrt{3}}{5}\right)\,,
\qquad
t_2 = \frac{1}{8\alpha_0} \ln\left(\frac{2+\sqrt{3}}{5}\right)\,.
\end{equation}%
That is, the $4D$ spacetime is accelerating in the interval
$1.4631>4|\alpha_0|t>0.1462$; during this interval the universe
expands by the factor of
\begin{eqnarray}
\frac{S(\tau_2)}{S(\tau_1)}=3.3810.
\end{eqnarray}
This is a small improvement over the decomposition $\R^{3+1}\times
\H_7$. It may be slightly further enhanced for smaller values of
$c$.

\subsection{$\R^{3+1}\times \H_{m_1}\times \H_{m_2}$ including radii}
\label{RHH}

In the previous sections, we have normalized $\e_i$ to $+1,\, 0$
or $-1$. Here we discuss the effect of including the radii
factors $r_i$ into the metric
\begin{equation}\label{hyperbolic3}
ds_{m}^2 = r_1^2 ds_{Hm_1}^2 + r_2^2 ds_{Hm_2}^2,
\end{equation}
where $r_1$ and $r_2$ are the physical curvature radii of
$\H_{m_1}$ and $\H_{m_2}$. Then the following $(4+m)$-dimensional
metric ansatz parametrized by the function $K=K(t)$:
\begin{equation}\label{mainsol}
ds_{4+m}^2 = {\rm e}^{2\lambda_0 t} \, ds_{R3}^2 + {\rm
e}^{-\frac{6\lambda_0 t}{m-1}} \left(- \, K^{\frac{2m}{m-1}} \,
dt^2 + K^{\frac2{m-1}} \, ds_{m}^2 \right)\,,
\end{equation}
solves the vacuum Einstein equations when
\begin{equation}\label{harmonic3}
K(t) = \frac{\lambda_0 r_1\,\gamma} {\sinh\left( \lambda_0 \beta
|t-t_1| \right)} \,,
\end{equation}
where
\begin{equation}
\beta = \sqrt{\frac{3(m+2)}{m}} \,, \quad %
\gamma = \sqrt{\frac{3(m+2)}{m(m-1)(m_1-1)}} \,, \quad %
r_1 = r_2 \, \sqrt{\frac{m_1-1}{m_2-1}}\,.
\end{equation}

The 4-dimensional Einstein metric is read off as \p{EinM1} with
the scale factor
\begin{eqnarray}
S(t)= e^{-\frac{3}{4}\lambda_0 t} K^{\frac7{12}},
\end{eqnarray}
for $m=7$. This is precisely the same scale factor one obtains for
a single hyperbolic space studied in~\cite{TW}, and there is not
much effect of introducing radii. We find that accelerated
expansion is again possible but it does not give enough expansion
factor.

\section{Hyperbolic external space}

In previous analysis of the class of time-dependent S-brane
solutions, the external space was taken to be flat.  In such cases
there was a period of accelerated expansion but the late time
limit of these solutions was decelerating.  In this section, we
consider the consequences of using a hyperbolic space instead of a
flat space as the large spatial dimensions.  We begin with a
review of the Milne solutions and vacuum solutions with hyperbolic
internal space.  Then we re-examine the vacuum solutions from
section~2, the difference being that we now go into Einstein frame
for the hyperbolic space. In this case we find that the period of
accelerated expansion vanishes but the late time behavior has
nearly constant expansion characterized by Milne spacetimes. Milne
spacetimes will later play a key role in our search for solutions
with eternally accelerating expansion.

\subsection{Milne spacetime limit}
\label{sec3.1}

Let us first consider the particular limit $\lambda_0 \to 0$.
Nontrivial solutions are obtained in this limit only for the
hyperbolic product space. (The topology of the whole spacetime is
$\R^{d+1}\times \H_{m_1}\times \cdots \times \H_{m_n}$.) Taking
this limit in eq.~\p{vs1}, we have $B_0=0$ and
\begin{eqnarray}
B_i &=& \frac{\bar \beta_i - \ln (t-t_1)}{m-1}, \nonumber\\
A &=& \frac{\bar \beta - m \ln (t-t_1)}{m-1},
\end{eqnarray}
where $m=\sum_{i=1}^{n} m_i$ and
\begin{eqnarray}
\bar \beta_i &=& \frac12 \ln \left[ \frac1{(m-1)(m_i-1)}
\prod_{j=1}^n \left( \frac{m_i-1}{m_j-1} \right)^{m_j}
\right],\nonumber\\
\bar \beta &=& \frac{m}2 \ln \left[
\frac1{m-1} \prod_{i=1}^n (m_i-1)^{-m_i/m} \right].
\end{eqnarray}
Therefore, the higher-dimensional vacuum solution, for $t_1=0$, is
\begin{equation}
\label{hdvs} ds^2 = - {\rm e}^{\frac{2\bar\beta}{m-1}} \;
t^{-\frac{2m}{m-1}} dt^2 + ds_{Rd}^2 + t^{-\frac2{m-1}}
\sum_{i=1}^n {\rm e}^{\frac{2\bar\beta_i}{m-1}} ds^2_{Hm_i},
\label{gmilne}
\end{equation}
This is a generalization of Milne metric.  For example when $n=1$,
the metric becomes
\begin{equation}
\label{milne} ds^2 = - d\xi^2 + ds_{Rd}^2 + \xi^2 ds^2_{Hm},
\end{equation}
where $\xi=[(m-1)t]^{-1/(m-1)}$. Ref.~\cite{Russo} discusses some
simple cosmological string models from Milne spacetime in four
dimensions, i.e., with $d=1$ and $m=2$.

We remind the reader that if we do not take quotients of the Milne
space, the above Milne spacetime can be obtained from a Wick
rotation of Euclidean space so these spacetimes are flat with all
Riemann curvature components vanishing.  In fact these solutions
cover just the patch of Minkowski space existing between the
lightcone of an observer. The hyperbolic space in this case is
just a result of a particular foliation of the spatial slices of
flat space. The maximal extension of a Milne universe is flat
Minkowski space which is not expanding. To understand this as an
expanding spacetime we must take a quotient of the hyperbolic
space as we discussed earlier.  (However since this quotient does
not affect the equations of motion, we leave the quotient
implicit here.)

\subsubsection{Flat space with hyperbolic extra dimensions}
\label{sec3.2}

Before examining the hyperbolic external space, let us consider
the hyperbolic internal space. We then find that the
Einstein-frame $(d+1)$-dimensional metric obtained from~\p{hdvs}
has the form of (\ref{EinM1}) with
\begin{equation}
S(t) = \exp \left[ \frac{\bar \beta - m \ln (t-t_1)}{(m-1)(d-1)}
\right] = \exp \left[ \frac{\bar \beta}{(m-1)(d-1)} \right] \; (t
- t_1)^{-\frac{m}{(d-1)(m-1)}}.
\end{equation}

We may redefine the time coordinate as ($t_1=0$)
\begin{equation}%
d\tau = - S^d(t) d t\,. \label{proper2}
\end{equation}%
We then have
\begin{equation}%
\tau = \frac{(d-1)(m-1)}{m+d-1} \exp\left[
\frac{d\bar\beta}{(d-1)(m-1)} \right]
t^{-\frac{m+d-1}{(d-1)(m-1)}},
\end{equation}%
so that the metric can be written in the standard FLRW form as
\begin{equation}
ds_{E(d+1)}^2 = - d\tau^2 + S^2(\tau) ds_{Rd}^2,
\end{equation}
with
\begin{equation}
S(\tau) = \exp\left( -\frac{\bar\beta}{m-d-1} \right) \left[
\frac{m+d-1}{(d-1)(m-1)}\,\tau \right]^{\frac{m}{m+d-1}}.
\end{equation}

The Hubble parameter is
\begin{equation}
H = \frac{\partial_\tau S}{S} = \frac{m}{m+d-1} \frac1{\tau}.
\end{equation}
Given that $d>1$, the value $m/(m+d-1)<1$ and hence
$\partial_\tau^2 S< 0$. Note that this result, dependent only on
the total dimension $m$ of internal spaces, is valid also for
Milne spacetime. We conclude that there is no inflation for the
above generalized Milne solutions~\p{gmilne} either. Next we turn
to the case where the usual $4D$ spacetime (i.e., the external
space) is hyperbolic.

\subsubsection{Interchanging external and internal spaces}
\label{interchange}

If we identify the large spatial $d$ dimensions as the hyperbolic
part of the spacetime given by (\ref{hdvs}), with the labels $d$
and $m$ interchanged, the induced metric on $\R \times \H_{d}$ in
the Einstein frame is given as \p{EinM1} with the scale factor
\begin{equation}
S(t) = \exp\left[ \frac{\bar\beta-\bar\beta_1}{(d-1)^2} \right]
\; t^{-\frac1{d-1}}.
\end{equation}
The internal space is then a flat $m$-dimensional space. As its
size is fixed, the internal space can be completely ignored, and
its dimension $m$ is irrelevant to our solution.

We may redefine the time coordinate as in \p{proper2} and then get
\begin{equation}
\tau = (d-1) \exp\left[\frac{d(\bar\beta-\bar\beta_1)}{(d-1)^2}
\right] t^{-\frac1{d-1}},
\end{equation}
so that the metric can be written in the standard FLRW form as
\begin{equation}
ds_{d+1}^2 = - d\tau^2 + S^2(\tau)\, d s_{H_d}^2,
\end{equation}
with
\begin{equation}
S(\tau) = \exp\left( -\frac{\bar\beta-\bar\beta_1}{d-1} \right)
\left( \frac{\tau}{d-1}\right). \label{hyper}
\end{equation}
These Milne spacetimes have linear expansion, and it is also clear
that the dimensional reduction of flat directions does not affect
this result. The Hubble parameter is
\begin{equation}
H = \frac{\partial_\tau S}{S} = \frac1{\tau}.
\end{equation}
This is the critical case with zero acceleration $\partial_\tau^2
S = 0$, and this fact will be used later.

Since hyperbolic internal spaces turn on positive effective
potentials in the effective $(d+1)$-dimensional theory, it is
natural to expect that if we turn the internal flat space into
hyperbolic spaces, we may increase the amount of inflation
generated. We study this possibility in more detail in subsection
~\ref{RHH2}.

\subsection{Hyperbolic space with flat extra dimensions}
\label{sec3.4}

Our exact solutions in subsection~\ref{higherdimgrav} can be
readily reinterpreted as solutions on hyperbolic (or spherical or
flat) external spaces with flat extra dimensions. Using
(\ref{vs1}), (\ref{ft1}) and (\ref{lambda}), and renaming some
variables, we rewrite the spacetime metric (\ref{metric}) as
\begin{eqnarray} \label{Sbranemetric}
ds^2 &=& -{\rm e}^{2 d g(t) - \frac{2m}{d-1} (\lambda_0 t +
\lambda_1)} dt^2 + {\rm e}^{2 g(t) - \frac{2m}{d-1} (\lambda_0 t
+ \lambda_1)} d\Sigma_{d,\s}^2 + {\rm e}^{2(\lambda_0 t +
\lambda_1)} ds_{Rm}^2
\nn \\
&=& {\rm e}^{-\,\frac{2m}{d-1}\,(\lambda_0 t+\lambda_1)} ds_E^2 +
{\rm e}^{2(\lambda_0 t+\lambda_1)} ds_{Rm}^2,
\end{eqnarray}
where the $(d+1)$-dimensional Einstein frame metric for the
external space is given by
\begin{equation}\label{hyperE}
ds_E^2 = - {\rm e}^{2 d g(t)} dt^2 + {\rm e}^{2g(t)}
d\Sigma_{d,\s}^2. \label{solh}
\end{equation}
Here $d$ is the dimension of the external space, which is flat,
spherical or hyperbolic for $\s = 0, 1, -1$, respectively. The
function $g(t)$ is given by
\begin{eqnarray} \label{gt}
g(t) &=& \left\{\begin{array}{ll}
\frac{1}{d-1} \ln \frac{\b}{\sinh[(d-1)\b(t-t_1)]} & :\s=-1, \\
\pm \b(t-t_1) & :\s=0, \\
\frac{1}{d-1} \ln \frac{\b}{\cosh[(d-1)\b(t-t_1)]} & :\s=+1,
\end{array}
\right. \\
\b &=& \frac{1}{d-1}\sqrt{\frac{m(m+d-1)}{d}} \; \lambda_0.
\end{eqnarray}
One can also derive the metric~(\ref{milne})
from~(\ref{Sbranemetric}) by taking the limit $\lambda_0\to 0$.
This is the exact solution with a critical expansion (zero
acceleration) discussed in the previous subsection. Here we
examine the solutions before discussing a limiting behavior, in
order to check whether we may get eternal or larger expansion.

The scale factor for the general exact solution \p{solh} is simply
$S(t)={\rm e}^{g(t)}$ with $g(t)$ given in \p{gt}. The conditions
for expansion and accelerated expansion are
\begin{eqnarray} \label{c1}
0&<&\frac{dS}{d\tau} = -\frac{\b}{S^{d-1}}\coth (d-1)\b t
\\
0&<& \frac{d^2S}{d\tau^2} = - \frac{(d-1)\b^2}{S^{2d-1}} \,.
\label{c2}
\end{eqnarray}
Obviously the first condition can be satisfied for $t<0$, but the
second condition never holds. There is no acceleration for this
exact solution. However, we can see that the Milne spacetime limit
$\b \to 0$ reproduces the critical expansion.

\subsection{Hyperbolic space with flat and hyperbolic extra dimensions}
\label{RHH2}

In the previous subsection we examined the amount of inflation we
could obtain when the internal dimensions were flat and the
external dimensions were hyperbolic.  Since hyperbolic spaces tend
to improve inflation, we here turn to the case where both the
internal and external spaces have hyperbolic dimensions. The
solutions obtained in subsection ~\ref{higherdimgrav} for the
product spaces $\R\times \H_{m_1}\times \R_d\times \H_{m_2}$ are
summarized in eqs.~(\ref{vs1}), (\ref{ft1}) and (\ref{metric}).
Instead of treating $\R_d$ as the external space, we now treat
$\H_{m_1}$ as the external space with $d$ and $m_1$ exchanged, and
the remaining flat and hyperbolic 
spaces are internal ones. Compared with the example in the
previous section, we have an additional hyperbolic space
$\H_{m_2}$ for the internal space. In this case, the Einstein
metric for the external space is
\begin{equation}
ds^2_{E,d+1} = {\rm e}^{\frac{2}{d-1}( dB_0 + m_2 B_2)} \left( -
e^{2A} dt^2 + e^{2B_1} ds^2_{Hd} \right),
\end{equation}
where the scale factor
\begin{equation}%
S = {\rm e}^{\frac{A-B_1}{d-1}} = \sinh^{\frac{-1}{d-1}} \lambda
(t-t_1) \ {\rm e}^{\frac{\b-\b_1}{(d - 1)(m-1)}}
\end{equation}%
can be simply written as
\begin{equation}
S =a_0  \sinh^x \lambda (t-t_1),
\end{equation}
where the constants are $x=-1/(d-1)$ and $a_0={\rm
e}^{\frac{\b-\b_1}{(d - 1)(m-1)}}$.  The key difference in taking
the external space to be hyperbolic instead of flat shows up in
the expansion factor $S$.  In this case the expansion factor is a
power of hyperbolic sine while in the other case the expansion
factor is a power of hyperbolic sine times an exponential
function.  We will find a significant difference in the behavior
of the universe.

Let us first examine the late time ($t\simeq 0_-$) and early time
($t\to -\infty$) behavior of the scale factor.  The late time
asymptotic behavior tells us whether there is eternally
accelerating expansion. To leading order, the scale factor close
to $t=0$ is
\begin{equation}
S \sim a_0 (\lambda t)^x
\end{equation}
where we have set $t_1=0$.  Surprisingly, different values of
$t_1$ does not seem to significantly change the results.  The
proper time is defined as $d\tau=-S^{d} dt \sim -\, a_0^{d}
(\lambda t)^{xd} dt$, so the relationship between the time $t$ and
the proper time $\tau$ is given by
\begin{equation}
\tau \sim -\,\frac{a_0^d (\lambda t)^{xd+1}}{xd+1} \sim (d-1)
a_0^d (\lambda t)^{xd+1} \quad \Rightarrow \quad t \sim
\tau^{-(d-1)}.
\end{equation}
Writing the scale factor in terms of the proper time we find
\begin{equation}
S\sim a_0 \lambda^x\,\tau^{-(d-1)x} \sim a_0\lambda^x\, \tau
\end{equation}
This solution has constant expansion but no eternally accelerating
expansion. It is a Milne type solution so it is closer to an
inflationary solution than when the external space was flat.

It turns out that the scale factor coincides with that in the
previous section with $\lambda=(d-1)\b$, and thus we see from
\p{c1} and \p{c2} that the exact solution is unfortunately always
decelerating. Using the above exact solution, we find that in the
$t \to 0$ limit and $d=3$, the deceleration to first order scales
as
\begin{equation}
\frac{d^2 S}{d\tau^2} \approx - \tau^{-5} \ . \label{hhfaccel}
\end{equation}
In the other limit $t\to -\infty$, the function $\sinh$ can be
approximated by the exponential function. It is easy to see in
this case that
\begin{equation}
S \propto \tau^{1/3},
\end{equation}
which is a dust filled universe not in the phase of inflation.

\subsection{Intersecting S-branes}

The solutions considered so far are all vacuum solutions. Here we
consider those with nonvanishing field strengths. In particular,
we examine some intersecting S-brane solutions. According to the
intersection rules~\cite{NO1,ADK}, we can construct an SM2-SM5
intersecting solution that leads to a 4-dimensional universe as
follows:
\begin{center}
\begin{tabular}{cccccccccccc}
& 1 & 2 & 3 & 4 & 5 & 6 & 7 & 8 & 9 & 10 & 0 \\
SM5 & $\bigcirc$ & $\bigcirc$ & $\bigcirc$ & $\bigcirc$ &
$\bigcirc$ &
$\bigcirc$ & & & & & \\
SM2 & $\bigcirc$ & $\bigcirc$ &  &  &  &  & $\bigcirc$ & & & &
\end{tabular}
\end{center}
where the world-volume directions of the S-branes are indicated
with a circle, and the remaining directions ($8,9,10,0$)
correspond to our spacetime. We find that the solution is given
by~\cite{NO1}
\begin{eqnarray}
ds_{11}^2 &=& K^{\frac13}L^{\frac23} \Big[ {\rm e}^{-2\gamma_1 t}
\left(-e^{6g(t)} dt^2 + {\rm e}^{2g(t)} d\Sigma_{3,\eps}^2 \right)
+ K^{-1} L^{-1} {\rm e}^{-2\gamma_1 t} (dy_1^2 + dy_2^2)
\nn\\
&& + L^{-1} \, {\rm e}^{\gamma_1 t}(dy_3^2 + \cdots + dy_6^2) +
K^{-1} \, {\rm e}^{4\gamma_1 t} dy_7^2 \Big]
\nn\\
&=& K^{\frac13} L^{\frac23} {\rm e}^{-2\gamma_1 t} ds_E^2 +
K^{-\frac23} L^{-\frac13} {\rm e}^{-2\gamma_1 t} (dy_1^2 + dy_2^2)
\nn\\
&& + K^{\frac13} L^{-\frac13} {\rm e}^{\gamma_1 t} (dy_3^2 +
\cdots + dy_6^2) + K^{-\frac23} L^{\frac23} {\rm e}^{4\gamma_1 t}
dy_7^2,
\end{eqnarray} %
where
\begin{equation}%
K(t) = \cosh(\c_2 \,t) \,, \qquad %
L(t) = \cosh(\c_3\,t) \,,
\end{equation}%
where $\c_1, \c_2$ and $\c_3$ are integration constants and
$\b=\sqrt{(18\c_1^2 +\c_2^2 + \c_3^2)/12}$ in $g(t)$. The metric
for our universe $ds_E^2$ coincides with vacuum solution~\p{solh}
with $g(t)$ given in \p{gt}. So unfortunately for this case we
find the same behavior as in subsection ~\ref{sec3.4}, and do not
get accelerated expansion. Actually it turns out that this
behavior is universal to all intersecting S-brane solutions as
long as we take the hyperbolic space to be the external space as
we show next.

\subsection{General intersecting solutions}

The general intersecting solution in $D$-dimensional supergravity
coupled to a dilaton and $n_A$-forms is given by~\cite{NO1}
\begin{eqnarray} \label{oursol} %
ds_D^2 &=& \prod_A [\cosh\tilde c_A (t-t_A)]^{2
\frac{q_A+1}{\Delta_A}} \Bigg[ {\rm e}^{2c_0 t+2c_0'} \left\{ -
{\rm e}^{2dg(t)} dt^2 + {\rm e}^{2g(t)} d\Sigma_{d,\eps}^2
\right\}
\nn\\
&& \hs{20} + \; \sum_{\a=1}^{p} \prod_A [\cosh\tilde
c_A(t-t_A)]^{-2\frac{\c_A^{(\a)}}{\Delta_A}} {\rm e}^{2 \tilde
c_\a t+2c_\a'} dx_\a^2 \Bigg],
\\
\phi &=& \sum_{A} \frac{(D-2)\e_A a_A}{\Delta_A} \ln \cosh\tilde
c_A(t-t_A) + \tilde c_\phi t + c_\phi',
\nn \\
E_A &=& \sqrt{\frac{2(d-2)}{\Delta_A}} \frac{{\rm e}^{\tilde
c_A(t-t_A) - \e_A a_A c_\phi'/2 + \sum_{\a\in q_A} c_\a'}}{\cosh
\tilde c_A(t-t_A)}, \quad %
\tilde c_A = \sum_{\a\in q_A} c_\a-\frac{1}{2} c_\phi \e_A a_A,
\\
&& c_0 = \sum_A \frac{q_A+1}{\Delta_A} \tilde c_A -
\frac{\sum_{\a=1}^p c_\a}{d-1}, \qquad %
c_0' = - \frac{\sum_{\a=1}^p c_\a'}{d-1},
\nn\\
&& \tilde c_\a = c_\a - \sum_A \frac{\d_{A}^{(\a)}}{\Delta_A}
\tilde c_A, \qquad %
\tilde c_\phi = c_\phi + \sum_A \frac{(d-2)\e_A a_A}{\Delta_A}
\tilde c_A, \nn\\
&& \frac{1}{d-1}\Big( \sum_{\a=1}^p c_\a \Big)^2 + \sum_{\a=1}^p
c_\a^2 +\frac{1}{2} c_\phi^2 = d(d-1) \b^2. \label{oursol1}
\end{eqnarray} %
where $D=d+1+p$, and $A$ denotes the kinds of $q_A$-branes, the
time derivative of $E_A$ gives the values of field strengths of
the antisymmetric tenors, $a_A$ is the parameter for the coupling
of dilaton and forms, and $\e_A= +1 (-1)$ corresponds to electric
(magnetic) fields.

Comparing eqs.~\p{oursol} and \p{einframe}, we find that the
prefactor in front of the $(d+1)$-dimensional line element is
given by
\begin{eqnarray} %
\prod_A [\cosh\tilde c_A (t-t_A)]^{2 \frac{q_A+1}{\Delta_A}} {\rm
e}^{2c_0 t + 2c_0'},
\end{eqnarray} %
and so the metric for our universe is again precisely given by
\p{hyperE}.  We conclude that this class of compactified theories
cannot give accelerated expansion except for the critical models
discussed in subsections~\ref{interchange} and \ref{RHH2}. This
conclusion is valid for the solutions to the vacuum Einstein
equations and also S-brane type solutions with nonvanishing field
strengths.

\section{Product spaces: effective action viewpoint}

\subsection{Dimensional reduction and effective action}

In subsection ~2.2 of \cite{CHNW}, and in \cite{LOW,EG}, pure
gravity in higher dimensions is rewritten as a lower dimensional
effective theory of gravity coupled to scalar fields by
dimensional reduction. The advantage of using the effective theory
formulation, compared with the higher dimensional gravity
viewpoint, is that we can directly deal with the
$(d+1)$-dimensional scale factor in the Einstein frame.
Furthermore, our experience in scalar field theories will be
useful in solving the equations of motion. For the reader's
convenience, we reproduce here some of the formulas that will be
used below.

The full spacetime is assumed to be a product space
$\R\times\M_0\times\cdots\M_n$. The ansatz for the metric is
\begin{equation}\label{metric-full}
ds^2 = \a^2 a^2(-d\eta^2 + d\Sigma_{d,\eps_0}^2) + \sum_{i=1}^n
a_i^2 d\Sigma_{m_i,\eps_i}^2,
\end{equation}
where $\eta$ is the conformal time in the $(d+1)$-dimensional
Einstein frame, $a$ is the $d$ dimensional scale factor and
\begin{equation}
\label{sum1} \a = \prod_{i=1}^{n} \a_i, \qquad \mbox{with} \qquad
\a_i = a_i^{-\frac{m_i}{d-1}}.
\end{equation}
Derivatives of $\eta$ will be denoted by a prime: $f'=df/d\eta$.

The dimensionally reduced theory of the Einstein gravity on this
space to the $(d+1)$ dimensional spacetime $\R\times\M_0$ is
given by $(d+1)$ dimensional gravity coupled to scalar fields
$\phi_i$ defined by
\begin{equation}
a_i = {\rm e}^{\phi_i}.
\end{equation}
The kinetic and potential terms for the scalar fields are
\begin{eqnarray}
&& \hs{-10} %
K = \frac{\rho+p}{2} = \sum_{i=1}^n
\frac{m_i(m_i+d-1)}{2(d-1)a^2} {\phi'_i}^2 + \sum_{i>j=1}^{n}
\frac{m_i m_j}{(d-1)a^2} \phi'_i \phi'_j - \eps_0
\frac{d-1}{2a^2}, \label{effK}
\\
&& \hs{-10} %
V = \frac{\rho-p}{2} = \sum_{i=1}^n (-\eps_i) \frac{m_i(m_i-1)}{2}
{\rm e}^{-\frac2{d-1} \left( (m_i+d-1) \phi_i + \sum_{j\neq
i}^{1\leq j\leq n} m_j \phi_j \right)} - \eps_0
\frac{(d-1)^2}{2a^2}. \label{effV}
\end{eqnarray}
The last terms in (\ref{effK}) and (\ref{effV}) are the
contributions from the curvature of the $d$ dimensional space.
{}From (\ref{effV}) we see that effective potentials arising from
gravity in higher dimensions are of exponential form. Recently,
the cosmology of multiple scalar fields with a cross-coupling
exponential potential was discussed in~\cite{GPCZ}.

The Einstein equations in the full spacetime are equivalent to
wave equations for each scalar field driven by these exponential
potentials plus the Friedman equation.

\subsection{Potential for $\M_1\times\M_2\times\M_3$}
\label{Potential}

Let us specialize to the product space of $(d+1)$-dimensional
universe and $\M_1\times\M_2\times\M_3$ with dimensions $m_1, m_2,
m_3$. Products of one or two spaces for extra dimensions can be
obtained by setting $m$'s to zero. The ansatz of the metric for
the full spacetime is
\begin{eqnarray} \label{metfors}
ds^2 = {\rm e}^{2\phi(x)} ds_{d+1}^2 + \sum_{i=1}^{3} {\rm
e}^{2\phi_i(x)} d\Sigma^2_{m_i,\epsilon_i},
\end{eqnarray}
where we have chosen the Einstein frame by setting
\begin{eqnarray} %
\phi= -\sum_i m_i \phi_i/(d-1).
\end{eqnarray}
The kinetic terms for the scalars corresponding to the radii of
each internal space in the effective theory are given by
(\ref{effK}). They can be diagonalized and normalized by a field
redefinition
\begin{eqnarray}
\psi_1 &=& \sqrt{\frac{m_1(m_1+d-1)}{d-1}} \Big[ \phi_1 +
\frac{1}{m_1+d-1} (m_2 \phi_2 + m_3 \phi_3) \Big],
\nn\\
\psi_2 &=& \sqrt{\frac{m_2(m_1+m_2+ d-1)}{m_1+d-1}} \Big[ \phi_2
+ \frac{m_3}{m_1+m_2+d-1} \phi_3 \Big],
\nn\\
\psi_3 &=& \sqrt{\frac{m_3(m_1+m_2+m_3+ d-1)}{m_1+m_2+d-1}}\phi_3,
\end{eqnarray}
with the result
\begin{eqnarray}
K &=& \frac{1}{2} \sum_{i=1}^3 \dot{\psi}_i^2 - \eps_0
\frac{d-1}{2a^2},
\\
V &=& - \sum_{i=1}^3 \s_i \frac{m_i(m_i-1)}{2} e^{\sum_a
M_{ia}\psi_a} - \e_0 \frac{(d-1)^2}{2a^2},
\end{eqnarray}
where the matrix $M_{ia}$ is given by
\begin{eqnarray}
M_{ia} = \left(\begin{array}{ccc}
-2\sqrt{\frac{m_1+d-1}{(d-1)m_1}} & 0 & 0 \\
-2\sqrt{\frac{m_1}{(d-1)(m_1+d-1)}} &
-2\sqrt{\frac{m_1+m_2+d-1}{m_2(m_1+d-1)}} & 0 \\
-2\sqrt{\frac{m_1}{(d-1)(m_1+d-1)}} &
-2\sqrt{\frac{m_2}{(m_1+d-1)(m_1+m_2+d-1)}} &
-2\sqrt{\frac{m_1+m_2+m_3+d-1}{m_3(m_1+m_2+d-1)}}
\end{array} \right).
\label{matrix}
\end{eqnarray}

To study the properties of the potential, it is more convenient
to define new independent fields as
\begin{eqnarray}
\vp_1 &\equiv& 2\sqrt{\frac{m_1+d-1}{m_1(d-1)}}\psi_1, \nn\\
\vp_2 &\equiv& 2\sqrt{\frac{m_1}{(d-1) (m_1+d-1)}} \psi_1+
2\sqrt{\frac{m_1+m_2+d-1}{m_2(m_1+d-1)}}\psi_2, \\
\vp_3 &\equiv& 2\sqrt{\frac{m_1} {(d-1)(m_1+d-1)}}\psi_1 +
2\sqrt{\frac{m_2}{(m_1+d-1)(m_1+m_2+d-1)}}\psi_2 \nn\\
&& + 2\sqrt{\frac{m_1+m_2+m_3+d-1}{m_3(m_1+m_2+d-1)}}\psi_3.\nn
\end{eqnarray}
The effective potential (\ref{effV}) is then
\begin{eqnarray}
V = - \sum_{i=1}^3 \s_i \frac{m_i(m_i-1)}{2} {\rm e}^{-\vp_i} -
\e_0 \frac{(d-1)^2}{2a^2}.
\end{eqnarray}
Clearly the potential is unbounded if any one of the $\s_i$'s is
positive. However, if we add contributions from antisymmetric
tensors, this is modified. For instance, the contribution of the
4-form field in 11-dimensional supergravity is
\begin{eqnarray}
\Delta V = b^2 \exp\Big[ -\frac{d(m_1 \vp_1 + m_2 \vp_2+m_3
\vp_3)} {m_1+m_2+m_3+d-1}\Big],
\end{eqnarray}
where $b$ is a constant. Now the potential is bounded from below
if $\frac{dm_i}{m_1+m_2+m_3+d-1}>1$ for the direction $\s_i=+1$
(there is no requirement in the direction $\s_i=-1,0$). There
will be local minimum in the direction with $\s_i=+1$, but the
potential minimum is always negative. This is basically the same
as what is discussed in \cite{EG} for one internal space, and it
seems difficult to get big expansion. In order to have big
expansion, we should have local minimum at positive value, where
our universe stays for a while and expands, and then decays to
lower value.

Nevertheless, it is already very interesting that a potential
with local minimum is obtained in the Einstein gravity coupled to
gauge fields, providing a mechanism for stabilizing the size of
extra dimensions. How to stabilize the size of extra dimensions
is an issue no less important than obtaining inflation. This is a
direction worthy of further exploration.

Another interesting research direction is to try to obtain
inflation in the present picture by introducing matter or
quintessence field into the solutions. This could be done, for
example, by considering D-branes in the solutions. We leave this
problem also to future study.

\subsection{Exact solutions for exponential potentials}
\label{exppot}

In this subsection we give a general discussion on solving scalar
field equations with exponential potentials in flat FLRW universe.
The same technique can be used to obtain solutions on hyperbolic
or spherical FLRW spacetime, as we will demonstrate in
section~\ref{eternal}.

Consider scalar fields coupled to gravity in $(d+1)$-dimensional
flat FLRW spacetime. The equations of motion for the scalar fields
are
\begin{eqnarray} \label{EOMd} %
\ddot{\psi}_a + d H \dot{\psi}_a + V_a = 0.
\end{eqnarray}
The Friedman equation is
\begin{eqnarray} \label{H2d}
\frac{d(d-1)}{2}H^2 = K + V, \qquad K \equiv \frac{1}{2}
\dot{\psi}_a^2.
\end{eqnarray}

For the potentials
\begin{equation}\label{V}
V = \sum_i v_i {\rm e}^{\sum_a M_{ia} \psi_a},
\end{equation}
we look for solutions of the form
\begin{equation}\label{ansatz1}
\psi_a = \a_a \ln t + \b_a, \qquad H = \frac{h}{t} \,.
\end{equation}
Eq.~(\ref{EOMd}) then implies that
\begin{eqnarray}
\sum_a M_{ia} \a_a &=& -2, \qquad \forall i, \nonumber \\
\Big( \frac{d(d-1)}{2} h - 1 \Big) \a_a + \sum_i u_i M_{ia} &=& 0,
\qquad u_i \equiv v_i {\rm e}^{\sum_b M_{ib} \b_b}.
\end{eqnarray}
These equations are not always consistent. To be precise, if
$M_{ia}$ is invertible, then $\a_a$ can be solved. From eq.
(\ref{H2d}) we find
\begin{eqnarray} \label{h}
h = \frac{1}{d-1} \sum_a \a_a^2.
\end{eqnarray}
If $M_{ai}$ is invertible, we have that
\begin{eqnarray} \label{al}
\a_a = -2 \sum_i M^{-1}_{ai},
\end{eqnarray}
\begin{eqnarray}
h = \frac{4}{d-1} \sum_{aij} M^{-1}_{ai} M^{-1}_{aj}.
\end{eqnarray}
There is (eternally) accelerating expansion if $h > 1$.

The results above can be easily generalized to nonstandard
kinetic term
\begin{equation}
K = \frac{1}{2} \sum_{A,B} g_{AB} \dot{\phi}_A \dot{\phi}_B.
\end{equation}
The potential term is still of the form (\ref{V})
\begin{equation}
V = \sum v_i e^{\sum_A M_{iA} \phi_A}.
\end{equation}
In this case we find that the expression for $h$ becomes
\begin{equation}
h = \frac{4}{d-1} \sum_{ijAB} M^{-1}_{Ai} g_{AB} M^{-1}_{Bj}.
\end{equation}

The ansatz (\ref{ansatz1}) allows us to find exact solutions for
product spaces with more than one independent scale factors, and
is a generalization of Milne spacetime discussed in section~3. In
the appendix we show how to obtain asymptotic solutions for
exponential potentials starting from the same ansatz
(\ref{ansatz1}).

\subsection{Check of acceleration}

The general solution in subsection ~\ref{exppot} provides an exact
solution for the spacetime in subsection ~\ref{Potential}. Let us
check if it has any acceleration phase. According to the
formula~\p{al}, we find from the matrix $M_{ia}$ in \p{matrix}
that
\begin{eqnarray}
\a_1 &=& \sqrt{\frac{(d-1)m_1}{m_1+d-1}}, \nn\\
\a_2 &=& \frac{(d-1)\sqrt{m_2}}{\sqrt{(m_1+d-1)(m_1+m_2+d-1)}},
\nn\\
\a_3 &=&
\frac{(d-1)\sqrt{m_3}}{\sqrt{(m_1+m_2+d-1)(m_1+m_2+m_3+d-1)}},
\end{eqnarray}
and the Hubble parameter $H=\frac{h}{t}$ is determined as
\begin{eqnarray}
h = \frac{1}{d-1} \a_a^2 = \frac{1}{2}\Bigg[ 1 -
\frac{d-1}{m_1+m_2+m_3+d-1} \Bigg].
\end{eqnarray}
This is always less than 1 for any $d$, in agreement with the
results in subsection ~\ref{sec3.2}.

\section{Eternally accelerating expansion on hyperbolic space}
\label{eternal}

We began our study of hyperbolic universe with hyperbolic extra
dimensions in subsection ~\ref{interchange}. In subsection
~\ref{RHH2} an exact solution was found but it did not lead to a
phase of accelerated expansion. In this section we examine the
critical solutions of subsection ~\ref{exppot} using the
dimensionally reduced scalar field theory formulation. In the next
subsection we will study the perturbations theory for this
critical solution.

Let the higher-dimensional geometry be $\R\times\H_d\times\H_m$.
The metric ansatz is
\begin{equation}%
ds^2 = {\rm e}^{-\frac{2m}{d-1}\,\phi(t)} \left( - dt^2 + a(t)^2
\, ds_{Hd}^2 \right) + {\rm e}^{2\phi(t)} ds_{Hm}^2 \,.
\end{equation}
The size of $\H_m$ corresponds to a scalar field in the effective
theory on $\R\times\H_d$. Here we write the (non-trivial)
components of the Einstein tensor for arbitrary $\epsilon_0$,
$\epsilon_1$:
\begin{eqnarray}
G_{00} &=& - \left[ \frac{\lambda}{2} \, \dot{\phi}^2 -
\epsilon_1 \frac{m(m-1)}{2} \, {\rm e}^{-\frac{2\lambda}{m}
\phi(t)} - \frac{d(d-1)}{2} \left( H^2 + \frac{\epsilon_0}{a^2}
\right) \right] \,, \label{G00}
\\
G_{xx} &=& - a^2 \left[ \frac{\lambda}{2}\, \dot{\phi}^2 +
\epsilon_1 \frac{m(m-1)}{2} \, {\rm e}^{-\frac{2\lambda}{m}
\phi(t)} + \frac{(d-1)(d-2)}{2} \left( H^2 +
\frac{\epsilon_0}{a^2} \right) + (d-1) \frac{\ddot{a}}{a}
\right], \label{Gxx}
\nn \\ \\
G_{ii} &=& \Bigg[ \frac{\lambda}{m} \left( \ddot{\phi} + d\,H
\dot{\phi} \right) - \frac{\lambda}{2} \dot{\phi}^2 - \epsilon_1
\frac{(m-1)(m-2)}{2} \, {\rm e}^{-\frac{2\lambda}{m}\phi(t)}
\nn \\
&{}&\qquad\qquad\qquad - \frac{d(d-1)}{2} \left( H^2 +
\frac{\epsilon_0}{a^2} \right) - d\,\frac{\ddot{a}}{a} \Bigg] {\rm
e}^{\frac{2\lambda}{m}\phi(t)}\,, \label{Gii}
\end{eqnarray} %
where $\lambda\equiv \frac{m(m+d-1)}{(d-1)}$. The metric on
$\R\times\H_d\times\H_m$ space takes the values
$\epsilon_0=\epsilon_1=-1$.

Henceforth we shall take $d=3$. Then by a change of variable
\begin{equation}
\phi = \sqrt{\frac{2}{m(m+2)}} \psi + \frac{1}{m+2}\ln(m(m-1)),
\end{equation}
we can simplify the Friedman equation and the wave equation for
$\psi$ as
\begin{eqnarray}
&3H^2 = \frac{1}{2} \dot{\psi}^2 + \frac{1}{2} {\rm e}^{-c\psi} +
\frac{3}{a^2}, \qquad c = \sqrt{\frac{2(m+2)}{m}}, \label{eqH}
\\
&\ddot{\psi} + 3 H \dot{\psi} - \frac{c}{2} {\rm e}^{-c\psi} = 0.
\label{eqpsi}
\end{eqnarray}
It is straightforward to obtain the critical solution with
$\ddot{a} = 0$
\begin{equation}\label{exactsol}
a = \sqrt{\frac{m+2}{2}} t, \qquad \psi = \frac{2}{c}\ln(t) +
\frac{1}{c}\ln\left( \frac{c^2}{8} \right).
\end{equation}

\subsection{Scalar perturbations} \label{perturb}

The exact solution with constant expansion is one of the solutions
we have focused on in this paper.  The main reason is that they
are the solutions which critically differentiate between
accelerating expansion and decelerating expansion.  Any tiny
perturbation should lead to interesting behavior and hopefully an
accelerating phase. We now turn to possible perturbations and show
how they may lead to solutions with eternally accelerating
expansion.

Let us consider a small perturbation around the
solution~(\ref{exactsol}). Let
\begin{equation}
a = a_0 + a_1, \qquad \psi = \psi_0 + \psi_1,
\end{equation}
where $a_0$ and $\psi_0$ are given by (\ref{exactsol}). It follows
that the Hubble parameter is
\begin{equation}
H = H_0 + H_1, \qquad H_0 = \frac{1}{t}, \qquad H_1 =
\frac{\dot{a}_1}{a_0} - H_0 \frac{a_1}{a_0},
\end{equation}
to the first order approximation. Expanding the equations
(\ref{eqH}) and (\ref{eqpsi}) and keeping first order terms only,
we get
\begin{eqnarray}
&6H_0 H_1 = \dot{\psi}_0 \dot{\psi}_1 - \frac{c}{2} {\rm
e}^{-c\psi_0} \psi_1 - 6 \frac{a_1}{a_0^3},
\\
&\ddot{\psi}_1 + 3 H_0 \dot{\psi}_1 + 3 H_1 \dot{\psi_0} +
\frac{c^2}{2} {\rm e}^{-c\psi_0} \psi_1 = 0 \,,
\end{eqnarray}
along with
\begin{equation}
2 \dot{\psi_0} \dot{\psi_1} + \frac{3\ddot{a_1}}{a_0} +
\frac{c}{2} \, {\rm e}^{-c\psi_0} \psi_1 = 0 \,.
\end{equation}
These linear equations are easy to solve. We find the following
solutions
\begin{equation}
\label{a1phi1} a_1 = A t^n, \qquad \psi_1 = \gamma A t^{n-1},
\end{equation}
where
\begin{equation}
\gamma = \frac{3(1-n)}{2\sqrt{m}} \,; \qquad n = \pm
\sqrt{\frac{m-6}{m+2}}\,.
\end{equation}
These give real solutions only if
\begin{equation}
m > 6\,.
\end{equation}
Note that $m=6$ or $n=0$ is excluded because it is just a zero
mode corresponding to time-shift symmetry. There are solutions
with eternally accelerating expansion when $m\geq 7$, although
$m=7$ is perhaps the most interesting case.  It is a very
intriguing numerical coincidence that $m=7$, which (together with
our spacetime 4 dimensions) is precisely the dimension in which
M-theory lives, is the minimum dimension for which this class of
perturbative solutions is allowed. This coincidence suggests that
the approach is worth serious consideration. The parameter $A$ is
not fixed, except that it has to be small so that the perturbative
expansion is valid.

For the case $m=7$, we have
\begin{equation}
n = \frac{1}{3}, \;\; -\frac{1}{3}; \qquad \gamma =
\frac{1}{\sqrt{7}}\,,\, \frac{2}{\sqrt{7}}\,.
\end{equation}
Examining the $n=1/3$ solution we find that
\begin{equation}
a_1 = A t^{1/3}, \qquad \ddot{a} = -\frac{2A}{9 t^{5/3}},
\end{equation}
so this gives positive acceleration for $A <0$. However, as time
increases, $a_1$ grows and perturbative expansion is no longer
valid. We can not claim eternally accelerating expansion for this
solution without further analysis. The other solution $n=-1/3$,
\begin{equation}
a_1 = A t^{-1/3}, \qquad \ddot{a} = \frac{4A}{9 t^{7/3}},
\end{equation}
gives a positive acceleration for $A > 0$. As time increases,
$a_1$ approaches to zero and our perturbative calculation remains
valid. We find eternally accelerating expansion for this case.

\begin{figure}[tp]
\begin{minipage}{70mm}
\begin{center}
\includegraphics[width=6cm]{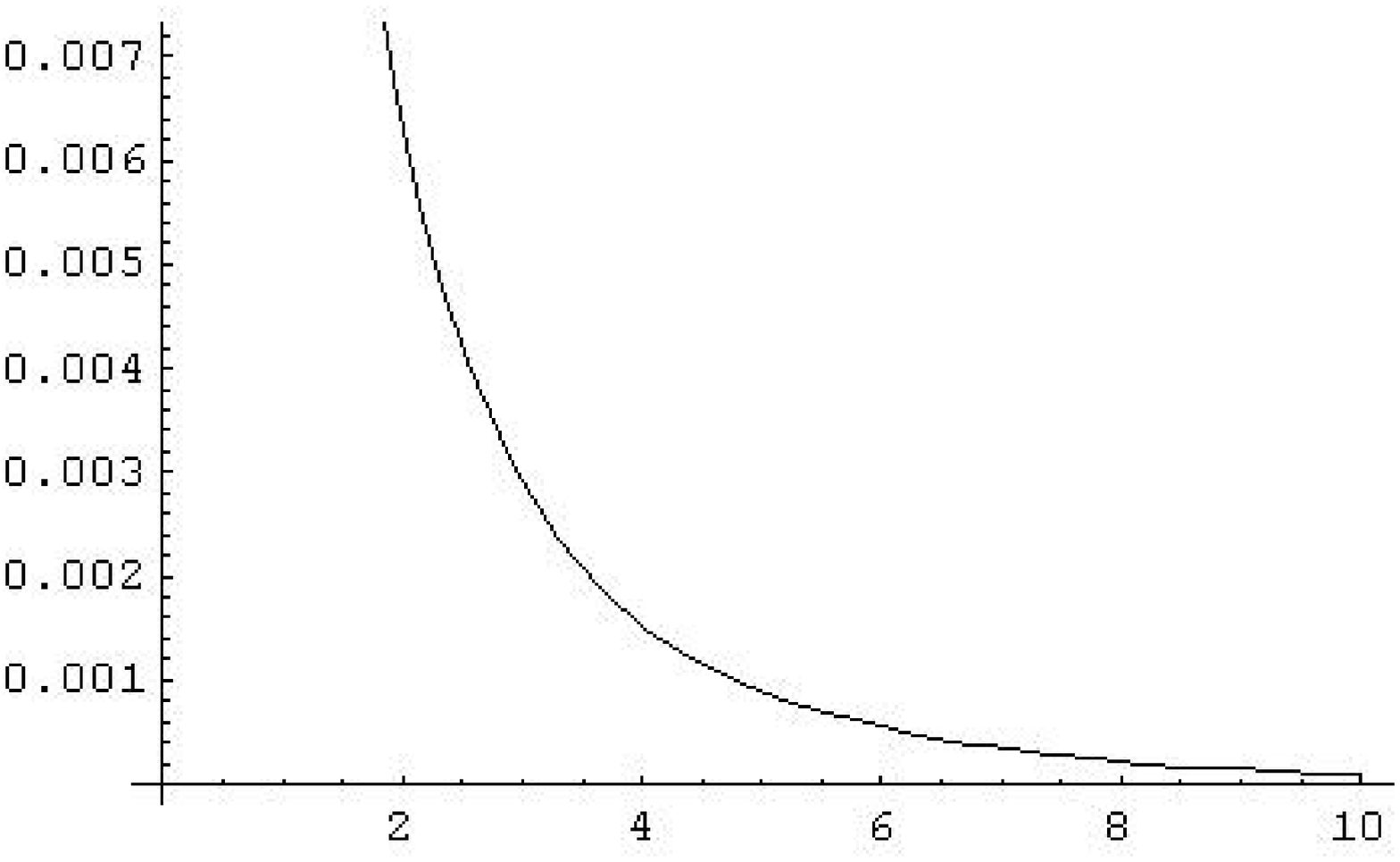}
\caption{ Acceleration $\ddot{a}$. } \label{HH-acc}
\end{center}
\end{minipage}
\hspace*{10mm}
\begin{minipage}{70mm}
\begin{center}
\includegraphics[width=6cm]{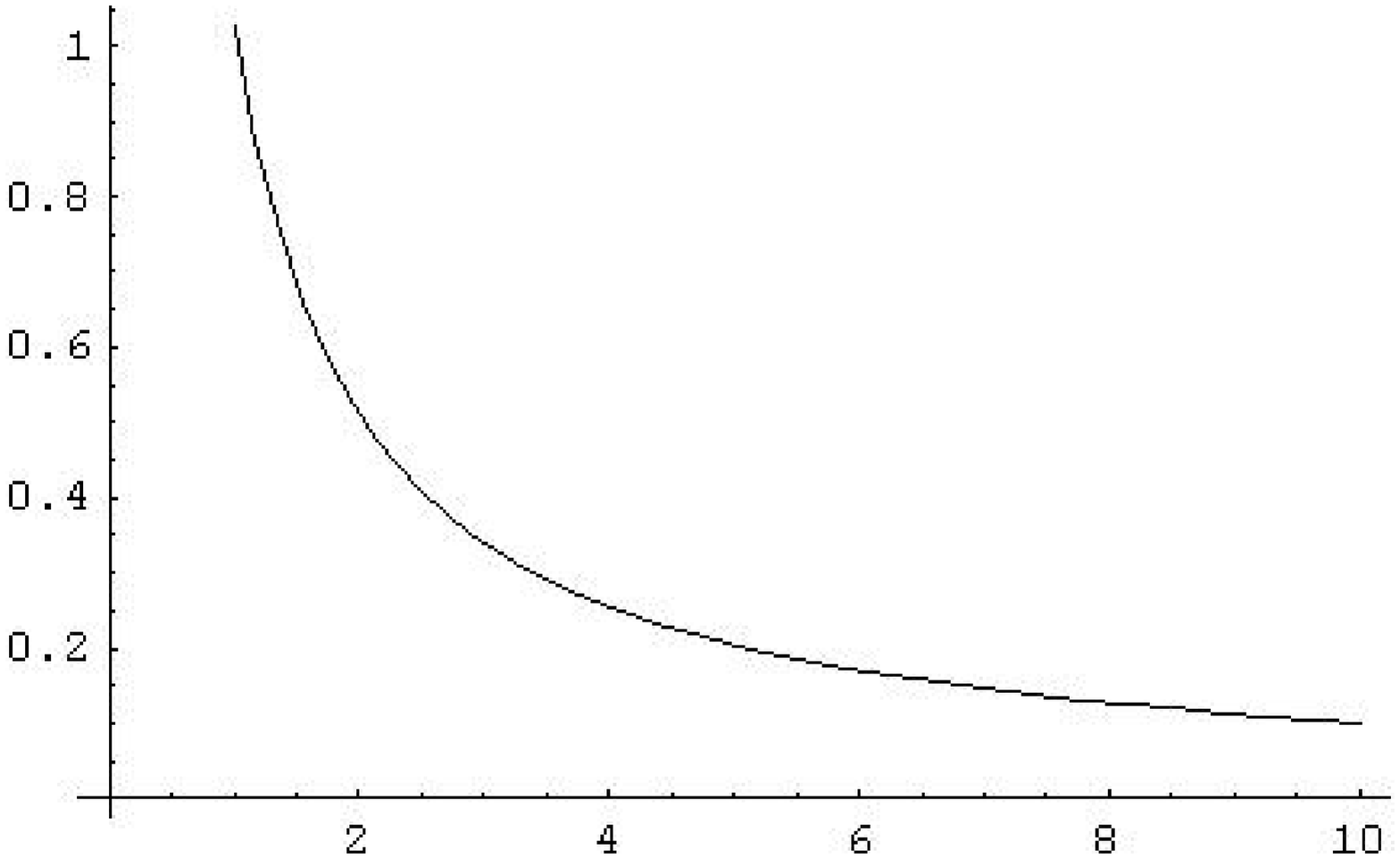}
\caption{ $\Delta a(t)$ } \label{HH-a-asym}
\end{center}
\end{minipage}
\hspace*{10mm}
\end{figure}

Numerical solutions can be explored to verify our claim of
eternally accelerating expansion without recourse to perturbation.
For the initial conditions
\begin{equation}
a=2.2, \qquad \phi=0.18, \qquad \dot{\phi}=0.2
\end{equation}
given at $t=1$, we find that the acceleration of the scale factor
$a$ is always positive but asymptotes to zero, as shown in
Fig.~\ref{HH-acc}. In Fig.~\ref{HH-a-asym}, the deviation of
$a(t)$ from the critical solution (\ref{exactsol}),
\begin{equation}
\Delta a \equiv\frac{a(t)-3t/\sqrt{2}}{3t/\sqrt{2}},
\end{equation}
is also shown to approach zero. This solution corresponds to $A
\simeq 0.01$ at $t=1$.

For product space compactifications, it is generally difficult to
find exact solutions for the coupled Einstein equations unless the
internal space is a product of flat spaces and at most one
nontrivial curved space or they all are of the same type.  It
would be interesting to find the exact solution corresponding to
the solution we have obtained here with eternally accelerating
expansion and see if the inflation is further increased at early
times and not just at late times.

\subsection{Turning on field strengths} \label{gaugefield}

In this subsection we would like to investigate the effects of a
non-zero form-field in a hyperbolic compactification by turning on
background field strengths, and study their effects on the $4D$
spacetime evolution. For simplicity we consider the full spacetime
to be the product of the large four-dimensions $\R \times \H_3$
and an extra hyperbolic space $\H_m$ of dimensions $m$. We will
focus more on this case because we have learned that an effective
gravity model where the usual $4D$ spacetime and internal spaces
both are hyperbolic may increase the amount of inflation.

In order to preserve the isotropy and homogeneity of the full
spacetime, the background field strength should respect the
isometry. For magnetic (electric) field background, the field
strength should be of the form
\begin{equation}\label{Fback}
F = f(t)\, \eps, \quad (F = f(t)\, dt\wedge\eps,),
\end{equation}
where $\eps$ is the volume form of either $\H_3$ or $\H_m$. Due
to electric magnetic duality, we only have to consider, say, the
electric and magnetic background (\ref{Fback}) with $\eps$ being
the volume form on $\H_3$.

\subsubsection{Electric field background}

Let us consider an example following \cite{Freund} where a field
strength is coupled to gravity. The model is the bosonic part of
11D supergravity, so we have a 3-form field $A$. The action is
just
\begin{equation}
S = \int d^{11} x \sqrt{-g}( R - \frac1{2\times 4!} F_{MNPQ}
F^{MNPQ}).
\end{equation}
We take the ansatz in the Einstein conformal frame \cite{Freund}
\begin{eqnarray}
&ds^2_{11} = {\rm e}^{-7\phi}(-dt^2 + a^2 ds^2_3) + e^{2\phi}
ds^2_7,
\\
& A_{abc} = \sqrt{g_3} \eps_{abc} A(t),
\end{eqnarray}
where $ds^2_3$ ($ds^2_7$) is for a 3-(7-)dimensional space, and
$g_3$ is the metric of $ds^2_3$. The Einstein equations are
\begin{equation}
G_{MN} = R_{MN} - \frac{1}{2} g_{MN} R = T_{MN},
\end{equation}
where
\begin{equation}
 T_{MN} = \left\{
\begin{array}{ll}
T_{\mu\nu} = - g_{\mu\nu} F^2(t), & \mbox{for} \;\; \mu,\nu =
0,1,2,3, \\
T_{mn} = g_{mn} F^2(t), & \mbox{for} \;\; m,n = 4,\cdots,10,
\end{array}
\right.
\end{equation}
where $F(t) \equiv {\rm e}^{14\phi}a^{-3}\dot{A}(t)/2$. The wave
equation for $F(t)$ is
\begin{equation}
\frac{d}{dt}({\rm e}^{7\phi}F(t)) = 0.
\end{equation}
This can be immediately solved as
\begin{equation}
F = f {\rm e}^{-7\phi},
\end{equation}
where $f$ is some constant.

An exact solution was given in \cite{Freund} with
\begin{equation}
a \propto t
\end{equation}
for $ds^2_3$ to be hyperbolic and $ds^2_7$ flat. This is the
critical case of zero acceleration. We can hope for the existence
of other solutions with acceleration, as we now show.

If $ds^2_3$ and $ds^2_7$ are both taken to be hyperbolic, the
11-dimensional field equations are
summarized by
\begin{eqnarray}
&3H^2 = \frac{63}{4} \dot{\phi}^2 + \frac{3}{a^2}
+ 21 {\rm e}^{-9\phi} + f^2 {\rm e}^{-21\phi}, \label{3H2} \\
&\ddot{\phi} + 3H \dot{\phi} - 6 {\rm e}^{-9\phi} - \frac{2}{3}
f^2 {\rm e}^{-21\phi} = 0. \label{phi2}
\end{eqnarray}

The terms proportional to $f^2$ in (\ref{3H2}) and (\ref{phi2})
are negligible compared to the other terms for large $t$. It is
easy to understand why this is the case, since the $f^2$ term is
proportional to ${\rm e}^{-21\phi}$, while the other potential
term is proportional to ${\rm e}^{-9\phi}$. (Note that $\phi$
increases with time in (\ref{exactsol}).)

We can analyze the effect of the electric background field by
treating it as a small perturbation, and repeating the same
technique utilized in section 5.1, for the zeroth order solution
(\ref{exactsol}). The only difference is that the electric field
serves as a source term for the first order perturbations of the
scale factor $a$ and the scalar field $\phi$. The solution is
again of the form (\ref{a1phi1}), but now with the amplitude $A$
fixed by the source term \be A = \frac{2^{11/6}f^2}{1215}, \ee and
\be n=-5/3. \ee  Once again we find a perturbation which leads to
acceleration for all time. As $t$ grows larger, the magnitudes of
the perturbations become smaller and the perturbative calculation
remains valid.
Note that, while the perturbation around the critical case can be
either accelerating or decelerating when there is no background
tensor field, there is only an accelerating perturbative solution
when we turn on the background field. In this sense the background
field assists the acceleration of the large dimensions.


\subsubsection{Magnetic field background}

The form field background is given by
\begin{equation}
F_{abc} = 2 f \sqrt{g_3} \, \eps_{abc},
\end{equation}
where $f$ is some constant, $\sqrt{g_3}$ is the (time independent)
unit volume of the three space, and $\eps$ is the (constant)
totally antisymmetrized tensor for directions $a,\,b,\,c = 1,2,3$,
and it vanishes for other (transverse) directions. The energy
momentum tensor due to the form field is \footnote{%
The convention of the form field appearing in the action is chosen
to be $-\frac1{2\times 3!} F_{MNP} F^{MNP}$.}
\begin{equation}
T_{00} = - \frac{g_3}{\tilde g_3} g_{00} f^2, \qquad T_{ab} =
\frac{g_3}{\tilde g_3} g_{ab} f^2, \qquad T_{mn} = -
\frac{g_3}{\tilde g_3} g_{mn} f^2,
\end{equation}
where $\tilde g_3$ is the determinant for the large spatial
dimensions $\H_3$ including the time dependent overall scale
factor.

For the ansatz
\begin{equation}
ds^2 = {\rm e}^{-m\phi}(-dt^2 + a^2 ds_{H3}^2) + {\rm e}^{2\phi}
ds_{Hm}^2,
\end{equation}
in which $\tilde g_3 = {\rm e}^{-3m\phi} a^6 g_3$, the Einstein
equations $G_{MN}=T_{MN}$ are equivalent to
\begin{eqnarray}
&\ddot{\phi} + 3H \dot{\phi} - (m-1)\, {\rm e}^{-(m+2)\phi} +
\frac{4{\rm e}^{2m\phi}}{(m+2)a^6} f^2 = 0, \\
&3H^2=\frac{m(m+2)}{4}\dot{\phi}^2 + \frac{3}{a^2} +
\frac{m(m-1)}{2}\, {\rm e}^{-(m+2)\phi} + \frac{{\rm
e}^{2m\phi}}{a^6} f^2\,.
\end{eqnarray}

In addition to the hyperbolic curvature term, $\phi$ has the
effective potential
\begin{equation}
V = \frac{(m-1)}{(m+2)} {\rm e}^{-(m+2)\phi} + \frac{2}{m(m+2)}
\frac{{\rm e}^{2m\phi}}{a^6} f^2,
\end{equation}
with a positive definite minimum for any given scale factor $a$,
and thus one may expect that eternally accelerating expansion is
possible in this case. However, just like the electric field
background, the form field effect can be ignored when $a$ gets
large. Therefore the asymptotic behavior should be the same as the
case without form field background.

Similar to the previous section on electric field background, we
can treat the effect of the magnetic field background by
perturbation theory. As an example, for $m=7$, the solution is
given by (\ref{a1phi1}) with
\begin{equation} A =
-\frac{3^{17/3}f^2}{2^{19/18}}, \quad n = \frac{1}{9}\,.
\end{equation}
This again gives a positive acceleration for the scale factor. In
this case the first order term of the scale factor grows with time
and the perturbative calculation can not be trusted for large $t$.
But solutions with eternally accelerating expansion should exist
because eventually the magnetic field can be ignored and we come
back to the case in section \ref{perturb}. Comparing this case
with the electric field background, we see that magnetic field
background has a more significant influence on the acceleration of
the universe.

\section{Conclusion}

In this paper we have studied the possibility of generating
inflation from gravity on product spaces with and without form
fields. For the only situation in which we obtained eternally
accelerating expansion in subsections
~\ref{perturb}-\ref{gaugefield}, the amount of inflation was
mainly boosted by the curvature of the hyperbolic external space,
when the initial condition is right. We give an explicit
description of the accelerating solution as a perturbation around
a critical case of constant expansion. When there is no
background form field, perturbation around the critical case can
be either accelerating or decelerating. A tensor field background
will increase the amount of acceleration, and we find that a
magnetic field in the large 3 dimensions (or equivalently the
electric field in extra dimensions) has a more significant effect
than the electric field in the large 3 dimensions (or magnetic
field in extra dimensions). All our other examples showed
inflationary phases with an e-folding number of order one.

A comment however is needed for the eternally accelerating
expansion we found. The number of e-foldings during an
inflationary phase is
\begin{equation}
\ln \frac{a(t_f)}{a(t_i)} = \int_{t_i}^{t_f} dt H(t),
\end{equation}
where $t_i$ and $t_f$ are the starting and ending time of the
inflationary phase. According to this definition, a large
e-folding number does not imply fast inflation. For our models
with eternally accelerating expansion in
subsections~\ref{perturb}-\ref{gaugefield}, the e-folding numbers
may be large mainly because although the inflation is small it
lasts forever, $
t_f\rightarrow
\infty$. In fact, since the curvature decreases when the space
expands, the curvature of the external space becomes negligible
within a short time, and the expansion of the universe is
negligibly better than constant expansion. This scenario is
apparently not suitable for cosmological inflation of the early
universe. As for the acceleration of the present universe,
we recall that, even for flat external space, the short
inflationary phase due to hyperbolic extra dimensions may be used
to explain the acceleration of the present universe with some
fine-tuning \cite{GKL}. Turning on negative curvature for the
large dimensions should help, but it might not play an important
role since our large spatial dimensions are known to be nearly
flat.

Our study suggests that a further generalization of the known
S-brane solutions would be desirable if we wish to use this model
for inflation. In addition to the obvious possibility of adding
positive scalar potentials by hand, or by quantum or stringy
corrections, pure gravity with warped or twisted geometry has not
yet been extensively studied. The possibility of introducing
matter (quintessence fields) by considering D-branes or other
extended objects ample in string/M-theory is another arena to
study. We save these topics for future study.

\section*{Acknowledgements}

The work of CMC, PMH, IPN, and JEW is supported in part by the
National Science Council, the Center for Theoretical Physics at
National Taiwan University, the National Center for Theoretical
Sciences, and the CosPA project of the Ministry of Education,
Taiwan, R.O.C. The work of PMH is also supported in part by the Wu
Ta-Yu Memorial Award of NSC. That of NO was supported in part by
Grants-in-Aid for Scientific Research Nos. 12640270 and 02041 from
Japan Society for the Promotion of Science.

\appendix

\section{Asymptotic solutions for exponential potentials}

In this appendix we demonstrate that the ansatz (\ref{ansatz1})
can be used to obtain asymptotic solutions for exponential
potentials.

As an example, consider the two field model with the kinetic and
potential terms
\begin{equation}
K = \frac{1}{2} (\dot{\psi}_1^2 + \dot{\psi}_2^2), \quad %
V = V_1 + V_2 = \eps_1 {\rm e}^{c\psi_1} + \eps_2 {\rm e}^{d\psi_1
+ f\psi_2}. \label{V2}
\end{equation}
Note that a more generic potential of the form
\begin{equation}
V = v_1 {\rm e}^{a\psi_1 + b\psi_2} + v_2 {\rm e}^{c\psi_1 +
d\psi_2}
\end{equation}
is equivalent to (\ref{V2}) by a rotation and translation of
$\psi_i$'s for $\eps_1 = {\rm sgn}(v_1), \eps_2 = {\rm sgn}(v_2)$.

The equations of motion are
\begin{eqnarray}
& \ddot{\psi}_1 + 3H \dot{\psi}_1 + cV_1 + dV_2 = 0, \label{eq1}
\\
& \ddot{\psi}_2 + 3H \dot{\psi}_2 + fV_2 = 0, \label{eq2}
\\
& 3 H^2 = K + V.
\end{eqnarray}
For large (cosmic) time $t$, we take the ansatz
\begin{equation}
\psi_1 = \a_1 \ln t + \b_1, \qquad \psi_2 = \a_2 t^{-n} + \b_2,
\quad H = \frac{h}{t},
\end{equation}
for $n > 0$. The idea is that the two equations of motion scale
differently for large $t$. Since $\psi_1$ decays slower than
$\psi_2$, we can have a nontrivial solution if $V_2$ decays
faster than $V_1$ and drops out of (\ref{eq1}), but remains
significant in (\ref{eq2}).

The solution is
\begin{equation}
\a_1 = -\frac{2}{c}, \quad n = \frac{2(d-c)}{c}, \quad h =
\frac{2}{c^2}, \quad u_1 = \frac{2(3h-1)}{c^2}, \quad u_2 =
\frac{1}{f}n(3h-n-1)\a_2,
\end{equation}
where
\begin{equation}
u_1 = e^{c\b_1}, \quad u_2 = e^{d\b_1 + f\b_2}.
\end{equation}
Note that $\a_2$ is not fixed by the field equations, but the
Hubble parameter is fixed.

The solution above is valid when the following conditions are met.
It is obvious that $n > 0$ only if $d > c$. The remaining
conditions are ${\rm sgn}(u_1) = \eps_1$ and ${\rm sgn}(u_2) =
\eps_2$.

Comparing the result with (\ref{h}), we see that the Hubble
parameter here looks as if the field $\psi_2$ is absent. It is
possible that a two-field model has eternally accelerating
expansion of this kind but not the kind given in the previous
section. To determine whether there is eternally accelerating
solution in this section, only the sign of $f$ is important; its
magnitude is not. But for the previous section the magnitude of
$f$ is also important.

\newcommand{\NP}[1]{Nucl.\ Phys.\ B\ {\bf #1}}
\newcommand{\PL}[1]{Phys.\ Lett.\ B\ {\bf #1}}
\newcommand{\CQG}[1]{Class.\ Quant.\ Grav.\ {\bf #1}}
\newcommand{\CMP}[1]{Comm.\ Math.\ Phys.\ {\bf #1}}
\newcommand{\IJMP}[1]{Int.\ Jour.\ Mod.\ Phys.\ {\bf #1}}
\newcommand{\JHEP}[1]{JHEP\ {\bf #1}}
\newcommand{\PR}[1]{Phys.\ Rev.\ D\ {\bf #1}}
\newcommand{\PRL}[1]{Phys.\ Rev.\ Lett.\ {\bf #1}}
\newcommand{\PRE}[1]{Phys.\ Rep.\ {\bf #1}}
\newcommand{\PTP}[1]{Prog.\ Theor.\ Phys.\ {\bf #1}}
\newcommand{\PTPS}[1]{Prog.\ Theor.\ Phys.\ Suppl.\ {\bf #1}}
\newcommand{\MPL}[1]{Mod.\ Phys.\ Lett.\ {\bf #1}}
\newcommand{\JP}[1]{Jour.\ Phys.\ {\bf #1}}



\begin{thebibliography}{99}
\itemsep 0pt

\bibitem{NM}
G. W. Gibbons, Aspects of supergravity theories, GIFT Seminar
1984, pp. 123-146, (QCD161:G2:1984), also in ``Supersymmetry,
 supergravity and related topics ''World Scientific, 1985;\\
J.~M.~Maldacena and C.~Nunez, ``Supergravity description of field
theories on curved manifolds and a no go theorem,'' Int.\ J.\
Mod.\ Phys.\ A {\bf 16} (2001) 822, arXiv:hep-th/0007018.

\bibitem{TW}
P.~K.~Townsend and M.~N.~Wohlfarth, ``Accelerating cosmologies
from compactification,'' Phys. Rev. Lett. {\bf 91} (2003) 061302,
arXiv:hep-th/0303097

\bibitem{NO2}
N.~Ohta, ``Accelerating cosmologies from S-branes,'' Phys. Rev.
Lett. {\bf 91} (2003) 061303, arXiv:hep-th/0303238.

\bibitem{CM}
C.~M.~Chen, D.~Gal'tsov and M. Gutperle, ``S-brane solutions in
supergravity theories,'' Phys. Rev. D {\bf 66} (2002) 024043,
arXiv:hep-th/0204071.

\bibitem{NO1}
N.~Ohta, ``Intersection rules for S-branes,'' Phys.\ Lett.\ B
{\bf 558} (2003) 213, arXiv:hep-th/0301095.

\bibitem{Roy}
S.~Roy, ``Accelerating cosmologies from M/String theory
compactifications,'' Phys. Lett. {\bf B} 567 (2003) 322,
arXiv:hep-th/0304084.

\bibitem{Wolf}
M.~N.~Wohlfarth, ``Accelerating cosmologies and a phase transition
in M-theory,'' Phys.\ Lett.\ B {\bf 563} (2003) 1,
arXiv:hep-th/0304089.

\bibitem{EG}
R.~Emparan and J.~Garriga, ``A note on accelerating cosmologies
from compactifications and S-branes,'' JHEP {\bf 0305} (2003) 028,
arXiv:hep-th/0304124.

\bibitem{NO3}
N. Ohta, ``A study of accelerating cosmologies from Superstring/M
theories,'' Prog. Theor. Phys. {\bf 110} (2003) 269,
arXiv:hep-th/0304172.

\bibitem{CHNW}
C.~M.~Chen, P.~M.~Ho, I.~P.~Neupane and J.~E.~Wang, ``A note on
acceleration from product space compactification,'' JHEP {\bf
0307} (2003) 017, arXiv:hep-th/0304177.

\bibitem{GKL}
M.~Gutperle, R.~Kallosh and A.~Linde, ``M / string theory,
S-branes and accelerating universe,'' JCAP {\bf 0307} (2003) 001,
arXiv:hep-th/0304225.

\bibitem{MI}
M. Ito, ``On the solutions to accelerating cosmologies,'' JCAP
{\bf 0309} (2003) 003, arXiv:hep-th/0305130.

\bibitem{LMP} H. L\"{u}, S. Mukherji, C. N. Pope and K.-W. Xu,
``Cosmological solutions in string theories,'' Phys. Rev. {\bf
D55} (1997)
7926, arXiv:hep-th/9610107;\\
H. L\"{u}, S. Mukherji and C. N. Pope, ``From $p$-branes to
cosmologies,'' Int. J. Mod. Phys. {\bf A14} (1999) 4121,
arXiv:hep-th/9612224.

\bibitem{LOW}
A.~Lukas, B. A. Ovrut and D.~Waldram, ``String and M theory
cosmological solutions with Ramond forms,'' Nucl. Phys. {\bf B495}
(1997) 365, arXiv:hep-th/9610238.

\bibitem{Cosmo}
L.~Cornalba and M.~S.~Costa, ``A new cosmological scenario in
string theory,'' Phys.Rev. {\bf D66} (2002) 066001,
arXiv:hep-th/0203031.

\bibitem{BQRSZ}
C. P. Burgess, F. Quevedo, S.-J. Rey, G. Tasinato and I. Zavala,
``Cosmological spacetimes from negative tension brane
backgrounds,'' JHEP {\bf 0210} (2002) 028, arXiv:hep-th/0207104

\bibitem{BNQTZ}
C. P. Burgess, C. N\'{u}\~{n}ez, F. Quevedo, G. Tasinato and I.
Zavala, ``General brane geometries from scalar potentials: gauged
supergravities and accelerating universes,'' JHEP {\bf 0308}
(2003) 056, arXiv:hep-th/0305211.

\bibitem{Russo}
J.~G.~Russo, ``Cosmological string models from Milne spaces and
$SL(2,Z)$ orbifold,'' arXiv:hep-th/0305032.

\bibitem{GPCZ} Z.-K. Guo, Y.-S. Piao, R.-G. Cai, and Y.-Z. Zhang,
``Cosmological scaling solutions and cross-coupling exponential
potential,'' arXiv:hep-th/0306245.

\bibitem{kaloper}
N.~Kaloper, J.~March-Russell, G.~D.~Starkman and M.~Trodden,
``Compact Hyperbolic Extra Dimensions: Branes, Kaluza-Klein Modes
and Cosmology,'' Phys. Rev. Lett. {\bf 85} (2000) 928,
arXiv:hep-ph/0002001.

\bibitem{silva}
S.~Nasri, P.~J.~Silva, G.~D.~Starkman and M.~Trodden, ``Radion
Stabiliation in Compact Hyperbolic Extra Dimensions,'' Phys.\
Rev.\ D {\bf 66} (2002) 045029, arXiv:hep-th/0201063.

\bibitem{ADK}
N. S. Deger and A. Kaya, ``Intersecting S-brane solutions of
$D=11$ supergravity,'' JHEP {\bf 0207} (2002) 038,
arXiv:hep-th/0206057.

\bibitem{Freund}
P.~G.~Freund, ``Kaluza-Klein Cosmologies,'' Nucl.\ Phys.\ B {\bf
209} (1982) 146.

\end{thebibliography}
\end{document}